\DeclareRobustCommand\openone{\leavevmode\hbox{\small1\normalsize\kern-.33em1}}

\newcommand{\oper}[2]{\hat{#1}^{\phantom{\dag}}_{\bf #2}}
\newcommand{\operdag}[2]{\hat{#1}^{\dag}_{\bf #2}}
\newcommand{\vecx}{\textbf{x}}
\newcommand{\vecy}{\textbf{y}}
\newcommand{\veci}{\textbf{i}}

\documentclass[aps,prl,twocolumn,groupedaddress,showpacs]{revtex4}

\usepackage{graphicx,amsmath,amssymb,color,psfrag}

\begin{document}

\title{FFLO- and N\'eel States in finite-size Systems}

\author{T. Gottwald}
\email{tobias.gottwald@uni-mainz.de}
\author{P.\ G.\ J. van Dongen}

\affiliation{KOMET 337, Institut f\"ur Physik, Johannes Gutenberg Univerit\"at, Mainz}

\date{\today}

\begin{abstract}
The general structure of the $s$-wave fermionic superfluid pairing order parameter is discussed for systems in thermal equilibrium. We demonstrate that for finite-size systems with fixed boundary conditions the pairing-amplitude may always be chosen as a {\it real} function in space, in contrast to systems underlying periodic boundary conditions, with drastical consequences for several postulated Fulde-Ferrell-Larkin-Ovchinnikov (FFLO) states. Using a simple mapping, we also investigate the consequences of our results for antiferromagnetic equilibrium states in a repulsive Hubbard model.
\end{abstract}


\maketitle

\section{1. Introduction}
Unconventional superfluid pairing such as Fulde-Ferrell-Larkin-Ovchinnikov (FFLO)-states have become of great interest recently, especially in the context of ultracold fermionic quantum gases \cite{koponen:finite:53dfgW}. As is well-known, BCS-, FFLO-, breached-pair-,  Sarma- and normal phases can occur in imbalanced two-component Fermi-mixtures, if there is an attractive interspecies $s$-wave interaction, which may be approximated as $U(\vecx - \textbf{y})= U \delta(\vecx - \textbf{y})$ in the low-temperature limit \cite{pethick:bose}. The particle density, population imbalance and temperature are the relevant parameters determining which phase is actually realized, either in model calculations or in nature \cite{koponen:finite:53dfgW}.

From a theoretical point of view, the key concept is the fermionic $s$-wave superfluid order parameter, which is generally described by a non-vanishing pair annihilation function $\Delta (\vecx , \vecx^\prime) \equiv \langle \oper{\psi}{\downarrow}(\vecx) \oper{\psi}{\uparrow}(\vecx^\prime) \rangle$. Here $\psi_\sigma (\vecx)$ destroys a fermion at position $\vecx$ in real-space and $\langle \ldots \rangle$ is the thermal average at inverse temperature $\beta \equiv 1/k_{\rm B} T$. In the context of Bose-Einstein condensates, the diagonal terms $\Delta(\vecx) \equiv \Delta(\vecx,\vecx^\prime = \vecx)$ are interpreted as a bosonic pair annihilation average at position $\vecx$ \cite{pethick:bose}.

The special feature associated with FFLO-states is a spatially varying order parameter, which is either considered to be a spatially varying complex function \cite{koponen:finite:53dfgW} or to have sign changes as a real function of $\vecx$ \cite{iskin:population:0FGw}. A common Ansatz for FFLO-states in translationally invariant systems is $\Delta (\vecx) = |\Delta| \, \exp(i\textbf{q} \cdot \vecx)$, where $\textbf{q}\neq {\bf 0}$ is the momentum carried by a Cooper pair in the FFLO-state \cite{koponen:finite:53dfgW,koponen:FFLO:gd78,silva:population:ok09,rombouts:unconventional:9ju,cui:polarized:pob4}. A BCS-state with $s$-wave symmetry in a balanced mixture would be described by $\textbf{q}={\bf 0}$.

FFLO-states may arise in experiments with superimposed optical lattices \cite{koponen:finite:53dfgW}, which would theoretically be described by Hubbard-type lattice models, as well as in continuous systems \cite{machida:generic:IIde8}. In this paper we discuss some general effects, arising from the finite system size, which is an immediate consequence of the presence of a trap potential in ultracold quantum gases. We show that a complex Ansatz for the superfluid order parameter is in general unnecessary in the presence of a confining potential, both for continuous and for discrete systems.

Another major experimental goal, apart from the search for FFLO-states, is presently the demonstration of antiferromagnetism in systems with superimposed optical lattices and repulsive interaction \cite{werner:interaction:5GJk7,koetsier:achieving:ko0,schneider:metallic:vfghi}. By mapping our results for superfluid states in the attractive Hubbard model to the repulsive-$U$ model, we show that the antiferromagnetic order parameter is aligned parallelly to a space-independent vector. This insight may significantly reduce the numerical effort in theoretical studies of antiferromagnetism in ultracold quantum gases.

Furthermore we present numerical results for superfluidity obtained in the saddle-point approximation \cite{andersen:magnetic:f890,iskin:population:0FGw,gottwald:antiferro:jhk6}. With these results we demonstrate, in particular, the importance of the Hartree terms in a trapped system, which in the literature are often neglected. We also present results for a Hubbard model with spin-dependent hopping and show that sign changes in the superfluid order parameter are not generally a feature of imbalanced systems, as seems to be the standard belief today.

This paper is organized as follows. First, in section 2, we discuss the structure of the superfluid order parameter in finite systems, depending on the boundary conditions. By a particle-hole transformation, these results are then, in section 3, related to antiferromagnetic states in the repulsive-$U$ Hubbard model. For the attractive-$U$ Hubbard model, we demonstrate in section 4 the importance of the Hartree terms in systems with non-vanishing magnetization. Excited states containing vortices are studied in section 5, and we present numerical results for spin-dependent hopping in section 6. Finally section 7 contains a summary and our conclusions.

\section{2. Structure of the order parameter}
In order to determine the structure of the superfluid order parameter, we first consider the macroscopic description of suprafluidity in continuum systems of finite extension and then discuss the consequences for finite discrete lattice systems.

It is standard knowledge that the macroscopic superfluid velocity and the superfluid current are described by the phase $\varphi (\vecx) \equiv \textrm{arg}[\Delta (\vecx)]$ of the order parameter $\Delta (\vecx)$ \cite{paananen:superfluid:2set6Z,leggett:quantum:bg09}. Specifically, the superfluid current is defined as ${\bf j}(\vecx) \equiv |\Delta (\vecx)|^2 \textbf{v}(\vecx)$, where $\textbf{v}(\vecx) \equiv \frac{\hbar}{m} \nabla \varphi (\vecx)$ is interpreted as the superfluid velocity. This implies that in FFLO states with a spatially varying complex superfluid order parameter a nonvanishing current occurs. We will now explain why such equilibrium solutions of FFLO form are suppressed in systems possessing a trapping potential, due to the finite size and the nature of the effective boundary conditions of such systems. 

While the continuity equation for the superfluid order parameter, discussed above, can readily be derived for superfluid {\it bosonic\/} systems, the derivation for fermionic systems is more difficult, and the result requires careful interpretation. For fermions, an additional term appears in the continuity equation, which arises from the kinetic term of the Hamiltonian, as follows:
\begin{equation}
 \frac{\partial}{\partial t} |\Delta(\vecx)|^2 + \nabla \cdot \textbf{j} = \frac{\hbar}{im} \Delta^{*} \, [\nabla_\vecx \cdot \nabla_{\vecx^\prime} \Delta(\vecx,\vecx^\prime)]_{\vecx^\prime \rightarrow \vecx}  + \textrm{h.c.}
\end{equation}
Since, apparently, both arguments $\vecx$ and $\vecx'$ are required independently to determine the right hand side of the continuity equation, it is clearly more convenient to consider a combined $2d$-dimensional variable $\vecy \equiv  (\vecx,\vecx^\prime)^T$ and focus on the full correlation function $\Delta(\vecx,\vecx^\prime)=\Delta(\vecy)$. One then arrives at a continuity equation of the following form:
\begin{equation}\label{vftOO0}
 \frac{\partial}{\partial t} |\Delta(\vecy)|^2 + \nabla \cdot \textbf{j}_S = \frac{1}{i \hbar} \Delta^* \left\langle \left[ \mathcal{H}_U , \hat{\Delta} (\vecy) \right] \right\rangle +\textrm{h.c.}  \; ,
\end{equation}
where we have defined the $2d$-dimensional ``supercurrent'' $\textbf{j}_S$ analogously to the BEC current, but now as a function of the generalized $2d$-dimensional coordinates $\vecy$. The right-hand side of \eqref{vftOO0} is the contribution from the interaction term and has the following explicit form for a contact interaction:
\begin{equation}\label{fg09OpP}
\begin{split}
  \frac{U \Delta^*}{i \hbar} \left\langle \operdag{\psi}{\uparrow} (\vecx) \oper{\psi}{\uparrow} (\vecx) \oper{\psi}{\uparrow} (\vecx^\prime) \oper{\psi}{\downarrow} (\vecx) \,-  \right.\hspace*{10ex} \\
\hspace*{10ex}\left. \operdag{\psi}{\downarrow} (\vecx^\prime) \oper{\psi}{\downarrow} (\vecx^\prime) \oper{\psi}{\downarrow} (\vecx) \oper{\psi}{\uparrow} (\vecx^\prime) \right\rangle  + \textrm{h.c.} 
\end{split}
\end{equation}
This contribution from the interaction term in the Hamiltonian is very small (negligible for our purposes) for various reasons. First, in our work we assume (as is customary for dilute BECs) that the interaction $U$ is {\it weak\/} \cite{pethick:bose}, so that this linear contribution in $U$ will tend to be numerically small.
Secondly, and more importantly, for superfluidity the diagonal terms $\Delta(\vecx,\vecx^\prime = \vecx)$ can be expected to be dominant because of the short-ranged $s$-wave interaction, and for $\vecx^{\prime} = \vecx$ the interaction term in \eqref{fg09OpP} vanishes {\it exactly\/}.
As a result one can to a very good approximation simplify \eqref{vftOO0} to 
\begin{equation}\label{vftOO00}
 \frac{\partial}{\partial t} |\Delta(\vecy)|^2 + \nabla \cdot \textbf{j}_S = 0\; .
\end{equation}
It will become clear below that \eqref{vftOO00}, which does not hold rigorously but is nevertheless very accurate, simplifies the structure of the equilibrium solutions drastically.

We now apply \eqref{vftOO00} to a condensate in equilibrium, i.e., to a time-independent low-temperature system {\it without\/} vortices. Accordingly we assume that the phase $\varphi (\vecy)$ of the order parameter has no singularities. From the literature we know that the curl of the superfluid velocity is generally quantized along a contour $C$ in a BEC \cite{pethick:bose}, i.e., 
\begin{equation}\label{bghII93}
 \frac{m}{\hbar} \oint_{C} d\textbf{s} \cdot \textbf{v}(\vecx) = 2 \pi n; \; n \in \mathbb{N} \; .
\end{equation}
For a condensate without any vortices ($n=0$), the superfluid velocity $\textbf{v}(\vecx)$ has no singularities and, hence, satisfies $\nabla \times \textbf{v}(\vecx)=0$ on account of \eqref{bghII93}. 
It is now easy to show from Gau\ss 's theorem that the superfluid order parameter $\Delta(\vecy)$ may be chosen to be {\it real\/}, provided that the boundary conditions are {\it fixed\/}, since then $\Delta(\vecy)=0$ and $\textbf{j}_S =0$ if $\vecx$ or $\vecx^\prime$ lies on the boundary:
\begin{eqnarray}\label{vfq789}
0 =\int_{\partial V} d{\bf F} \cdot (\varphi \, \textbf{j}_S) &=& \int_{V} dV \, \nabla \cdot \left( \varphi \, \textbf{j}_S \right)
\\ \nonumber
 \approx \int_{V} dV \, \nabla  \varphi \cdot \textbf{j}_S  &=& \frac{m}{\hbar} \int_{V} dV \left| \frac{\textbf{j}_S}{\Delta(\vecy)} \right|^2 ,
\end{eqnarray}
In the derivation of \eqref{vfq789} we have used \eqref{vftOO0} or \eqref{vftOO00}, i.e., $\nabla \cdot \textbf{j}_S =0 $. It now follows from \eqref{vfq789} that $\textbf{j}_S=0$ for all values of $\vecy$, and we may, therefore, choose $\Delta(\vecx,\vecx^\prime) \in \mathbb{R}$ for all $\{\vecx, \vecx^\prime\}$. For periodic boundary conditions this argument fails, since the quantities mentioned above do not vanish at the boundaries.

\subsection{Lattice systems}
The properties of the superfluid order parameter, defined for smooth coordinates $\vecx$, may be mapped onto its discrete analogue $\Delta (\veci)$, defined on a lattice, by the use of wave functions describing localized states $w_{\veci} (\vecx)$ at lattice positions $\veci$ \cite{paananen:noise:op0e}; if the trapping potential is switched off, the states $w_{\veci} (\vecx)$ would be the Wannier-functions in the lowest band. When an optical lattice is superimposed, the quantum gas is effectively described by a discretized second-quantized Hamiltonian of Hubbard-type form \cite{iskin:population:0FGw},
\begin{eqnarray}\label{chT6a}
\mathcal{H} &=& -t \sum_{( {\bf ij} ), \sigma} \operdag{c}{i \sigma} \oper{c}{j \sigma} + \sum_{\bf i \sigma} \left( V{\bf i}^{2} - \mu_{\sigma} \right) \oper{n}{i \sigma} \\
\nonumber &+& U \sum_{\bf i} \oper{n}{i \uparrow} \oper{n}{i \downarrow} \; .
\end{eqnarray}
Here $t$ is the nearest-neighbor hopping, $V>0$ describes the steepness of the parabolic trap, the parameters $\mu_\sigma$ (with $\sigma =\uparrow ,\downarrow$) describe the on-site potentials, and $U$ is the on-site interaction strength, which may be chosen to be attractive \cite{zwierlein:fermionic} or repulsive \cite{schneider:metallic:vfghi}. The center of the harmonic trap is located at $\veci=0$.

Our arguments imply that, also in this Hubbard-type model, the superfluid order parameter $\Delta(\veci)$ may always be chosen as a real function (provided, of course, that $U$ is attractive), since due to the confining potential $V_\sigma$ there is no occupation for large $|\veci|$, and therefore no superfluid current is present at the ``border'' of the system. Here we have assumed that $w_{\veci} (\vecx) \in \mathbb{R}$, which is {\em exact} for $V=0$ and should therefore be quite accurate also in the general case.

Iskin and Williams \cite{iskin:population:0FGw} and Chen et al. \cite{chen:exploring:kl4dw} assumed $\Delta (\veci)$ to be {\it real\/} for numerical simplicity. Our arguments show that there is, in fact, no need for a complex superfluid order parameter, if vortices are excluded. On the other hand there is a multitude of literature \cite{koponen:finite:53dfgW,silva:population:ok09,rombouts:unconventional:9ju,cui:polarized:pob4,koponen:FFLO:gd78} using various Ansatzes with complex forms of $\Delta (\veci)$ for trapped quantum gases, especially in the context of the local density approximation (LDA). These results are not consistent with our arguments and should therefore be carefully rechecked.
\begin{figure*}
 \begin{minipage}{0.66\columnwidth}
  \resizebox{1.0\columnwidth}{!}{\includegraphics[clip=true]{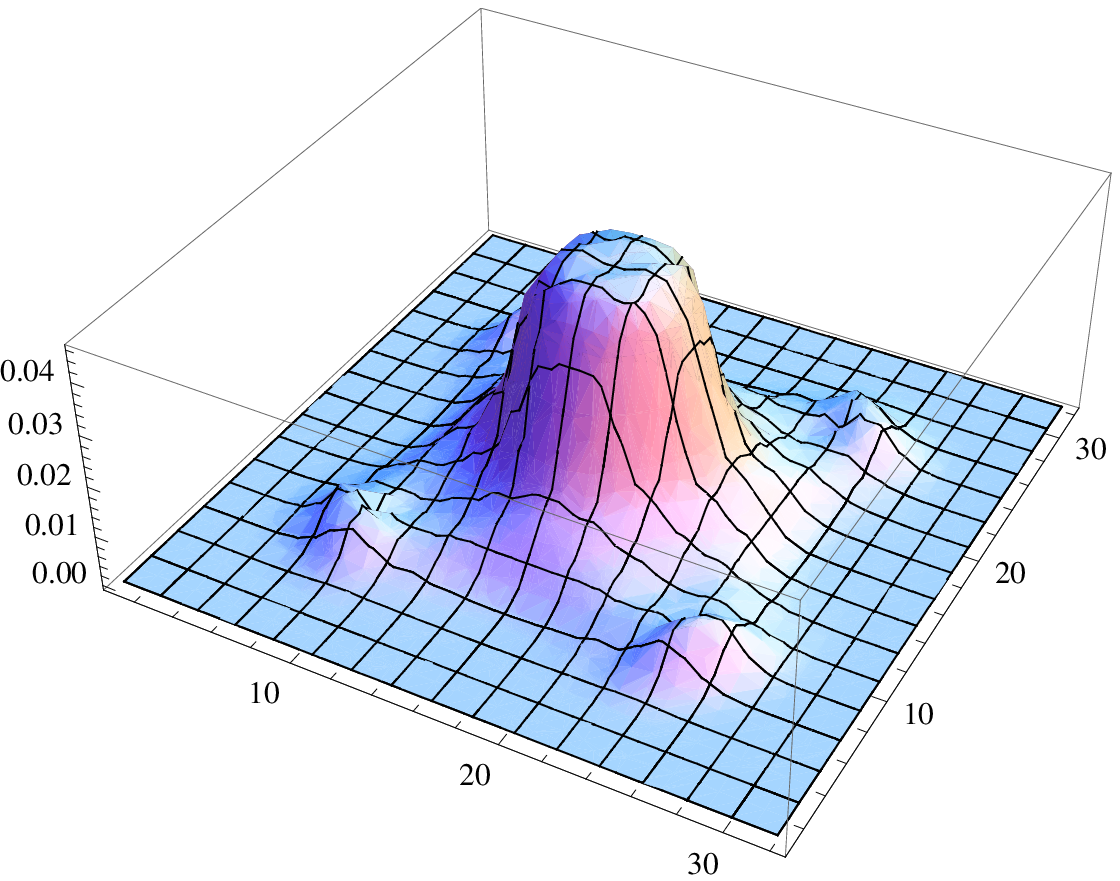}}
 (a)
 \end{minipage}
 \begin{minipage}{0.66\columnwidth}
  \resizebox{1.0\columnwidth}{!}{\includegraphics[clip=true]{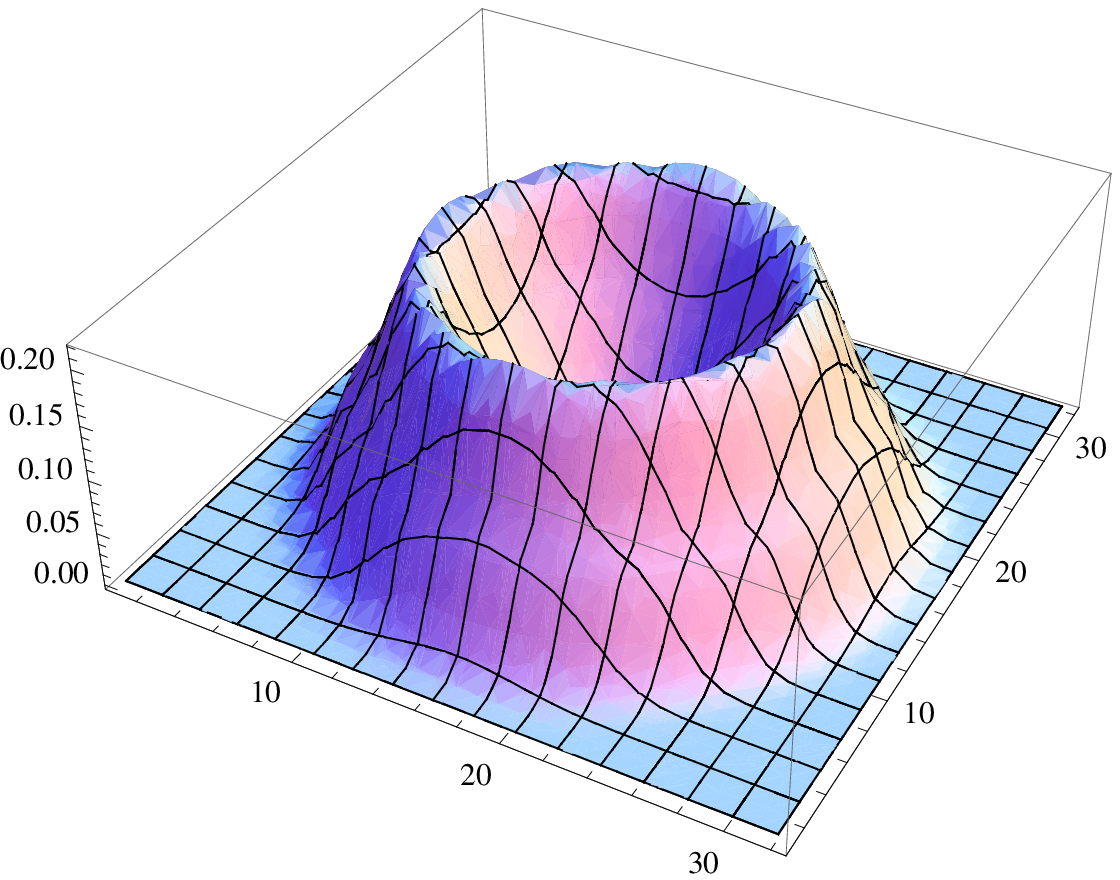}}
 (b)
 \end{minipage}
 \begin{minipage}{0.66\columnwidth}
  \resizebox{1.0\columnwidth}{!}{\includegraphics[clip=true]{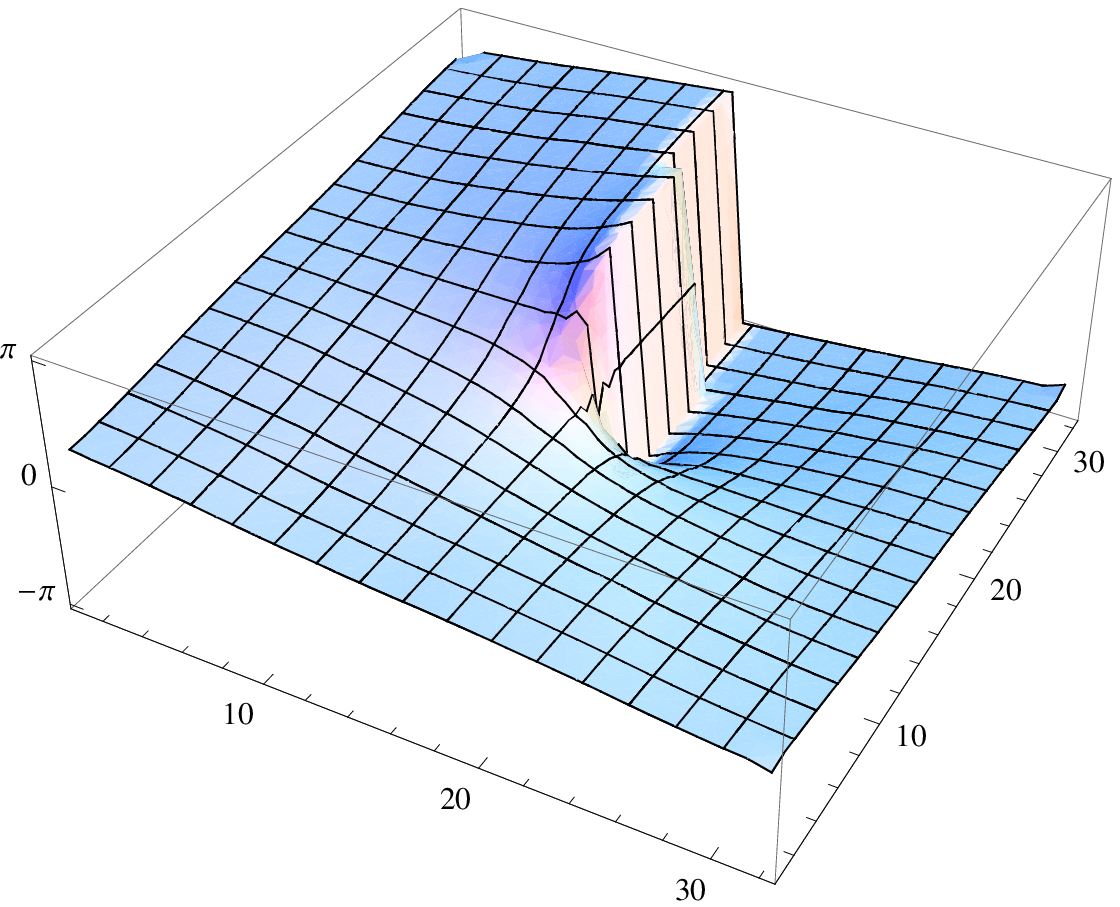}}
 (c)
 \end{minipage}
\caption{Magnetization (a), $|\Delta(\veci)|$ (b) and $\varphi(\veci)$ (c) in a vortex state with $n=1$ depending on $\veci$-position on a $32 \times 32$ lattice. While $|\Delta(\veci)|$ depends only on the radial position, the magnetization is maximal in the trap center, where $\varphi$ is singular. In contrast to a vortex-free system, the magnetization profile does not break the $\pi/2$-rotational symmetry of the lattice. The system parameters have been chosen as: $U=-2.5$, $V=0.025$, $\beta=1000$, $\mu \equiv (\mu_\uparrow + \mu_\downarrow)/2=0.5$ and $\Delta \mu =0.4$ in units of $t$.}\label{gt66ui0}
\end{figure*}

Since these analytical results concerning the general structure of the order parameter rely on approximative assumptions, we emphasize that, for the case of the Hubbard model \eqref{chT6a}, these results may also be confirmed numerically (within the unrestricted Hartree-Fock approximation). Starting from periodic boundary conditions one may obtain complex (vortex-free) solutions of the order parameter $\Delta(\veci)$ by using $\Delta_{0} (\veci) = \Delta \exp(i \textbf{q} \cdot \veci)$ as the starting point for a numerical iteration. After convergence of the iteration process is reached, the sum over the squares of the imaginary parts $\sum_{\veci} \Im\{\Delta(\veci)\}^2$ may be minimized by a global phase rotation as follows: $\Delta(\veci) \rightarrow \Delta(\veci) \exp(i \varphi) \; \forall \, \veci$. For periodic boundary conditions one finds (after the minimization): $[\sum_{\veci} \Im\{\Delta(\veci)\}^2]/[\sum_{\veci} |\Delta(\veci)|^2] \approx 0.3$. If one reduces the hopping amplitudes across the periodic boundaries to a relative strength of 20\% one obtains $[\sum_{\veci} \Im\{\Delta(\veci)\}^2]/[\sum_{\veci} |\Delta(\veci)|^2] \approx 0.008$ and if one turns the hopping amplitudes across the periodic boundaries off (fixed boundary conditions) the imaginary part reduces to $[\sum_{\veci} \Im\{\Delta(\veci)\}^2]/[\sum_{\veci} |\Delta(\veci)|^2] \approx 10^{-5}$, with otherwise the same parameters. This is in perfect agreement with our previous analytical statement: Also according to our numerical results, complex order parameters (without vortices) may occur for periodic boundary conditions, while, for fixed boundaries, the order parameter can be chosen as a {\em real} function, $\Delta (\veci) \in \mathbb{R}$.

\section{3. Connection to the repulsive-$U$ model}
The Hubbard model for attractive interaction is connected to the repulsive model (for bipartite lattices) via a canonical (so-called special particle-hole) transformation \cite{liebWu:HubbardModel}:
\begin{equation}
 \oper{c}{i \uparrow} \rightarrow (-1)^{\veci} \operdag{c}{i \uparrow} \quad , \quad \oper{c}{i \downarrow} \rightarrow \oper{c}{i \downarrow} \; ,
\end{equation}
where $(-1)^{\veci}$ is $+1$ on the $A$-sublattice and $-1$ on the $B$-sublattice. Under this canonical transformation, the Hubbard Hamiltonian \eqref{chT6a} is transformed into
\begin{eqnarray}\label{durb8H}
\mathcal{H} &\rightarrow& -t \sum_{( {\bf ij} ), \sigma}  \operdag{c}{i \sigma} \oper{c}{j \sigma} - \sum_{\bf i \sigma} \left( \sigma V{\bf i}^{2} + \mu^{\prime}_{\sigma} \right) \oper{n}{i \sigma} \\
\nonumber &-& U \sum_{\bf i} \oper{n}{i \uparrow} \oper{n}{i \downarrow} +\textrm{const} \; ,
\end{eqnarray}
while simultaneously the superfluid order parameter $\Delta (\veci) = \langle \oper{c}{i \uparrow} \oper{c}{i \downarrow} \rangle$ is transformed into a staggered magnetic order parameter in the $xy$-plane (and vice versa):
\begin{equation}
\langle \oper{c}{i \uparrow} \oper{c}{i \downarrow} \rangle \leftrightarrow (-1)^{\veci} \langle \operdag{c}{i \uparrow} \oper{c}{i \downarrow} \rangle = (-1)^{\veci} (S_{\veci, x} + i S_{\veci, y}) \; .
\end{equation}
Hence, the Hubbard Hamiltonian \eqref{chT6a} with a repulsive $U$ is transformed into a Hamiltonian with an attractive $U$ and a spatially varying Zeeman-term. On account of this Zeeman term, any equilibrium state of \eqref{durb8H} has a vanishing occupation for the ``down''-spin species ($n_{\veci \downarrow} \simeq 0$) for large $|\veci|$ and a full occupation for the ``up''-spin species ($n_{\veci \uparrow} \simeq 1$). As a consequence, in the negative-$U$ model, the superfluid flow vanishes in the large-$|\veci|$ regime. Hence, the superfluid order parameter $\Delta(\veci)$ may globally be chosen to be {\it real\/} for $U<0$ on the basis of the arguments presented in section 2. Transforming this {\it real\/} order parameter for $U<0$ back to the repulsive-$U$ model demonstrates that, in this case, the $xy$-antiferromagnetic order may always be chosen along the $x$-direction, confirming the numerical results found in \cite{gottwald:antiferro:jhk6} within Hartree-Fock calculations  and the results found in \cite{PhysRevB.83.054419} within R-DMFT calculations for particle-imbalanced Fermi-mixtures.

In balanced systems the symmetry of the underlying Hubbard model is larger\cite{gottwald:antiferro:jhk6}, namely $SU(2)$ instead of $U(1)$, and, therefore, antiferromagnetic order is allowed also along the $z$-direction. For a balanced system, our arguments imply that the order parameter has a unique (globally fixed) direction. For instance a spiral structure in the $xz$-plane cannot occur, since, for such a state, one can rotate the system around the $x$-axis [since the balanced Hamiltonian is fully $SU(2)$-invariant] to obtain an incommensurate $xy$-antiferromagnet, which is not allowed on account of our arguments presented above. Hence, also for balanced systems, the direction of the antiferromagnetic order parameter is globally defined. This result depends, of course, critically on the presence of a trap, which determines the boundary conditions. Indeed, in translationally invariant systems away from half-filling, incommensurate states are well-known and have been predicted analytically long ago \cite{schulz:incommensurate:bg89I}.

\section{4. Role of the Hartree terms}
In sections 5 and 6 (see below) we will illustrate the above ideas and present numerical results for the attractive-$U$ Hubbard model in the saddle-point approximation, which takes in general the well-known form
\begin{eqnarray}\label{vGt2ws}
 \oper{n}{i \uparrow} \oper{n}{i \downarrow} &\rightarrow&  \Delta(\veci) \left( \oper{c}{i \uparrow} \oper{c}{i \downarrow} + \operdag{c}{i \downarrow} \operdag{c}{i \uparrow} \right) - \Delta^2(\veci) \\
\nonumber &+& \langle \oper{n}{i \uparrow} \rangle \oper{n}{i \downarrow} + \oper{n}{i \uparrow} \langle \oper{n}{i \downarrow} \rangle - \langle \oper{n}{i \uparrow} \rangle \langle \oper{n}{i \downarrow} \rangle 
\end{eqnarray}
and may require the following comment. In several previous publications \cite{iskin:population:0FGw,chen:exploring:kl4dw}, the {\it Hartree terms\/}, which correspond to the second line in Eq.\ \eqref{vGt2ws}, were neglected for numerical simplicity, since they were assumed to constitute an irrelevant deformation of the trapping potential $V\veci^2$. This assumption, however, is only correct in spatial regions where the magnetization vanishes ($m_\veci \equiv n_\uparrow - n_\downarrow = 0$). In magnetized spatial regions the Hartree terms create an additional effective space-dependent Zeeman-term, which leads to a decrease of the local magnetization for $U<0$. As a consequence, for the purposes of this paper, we must include the Hartree terms in our calculations, since they are of the same order of magnitude as the superfluidity terms [first line in Eq.\ \eqref{vGt2ws}]. In fact, we find that inclusion of the Hartree terms is physically highly significant and leads to results {\it differing\/} from previous literature \cite{chen:exploring:kl4dw}, in which they were neglected. Specifically, if Hartree terms are neglected one finds parameter regions, in which the superfluid order parameter has sign changes in the tangential direction, while the magnetization displays a rapidly oscillating behavior, as visible, e.g., in Fig.\ 4(g) in Ref.\ \cite{chen:exploring:kl4dw}. However, inclusion of the Hartree terms leads to a simpler radial oscillation, similar to the behavior visible, e.g., in Fig.\ 3(e) in Ref.\  \cite{chen:exploring:kl4dw}. In \cite{chen:exploring:kl4dw} it is shown that, if Hartree terms are neglected, the structure of the order parameter strongly depends on the filling in the center of the trap $n_C$. Including the Hartree terms we find the following results as compared to Ref.\ \cite{chen:exploring:kl4dw}:
\begin{enumerate}
 \item In the low-filling region ($n_C<1$) we find qualitatively the same results.
 \item In the high-filling regime ($n_C \approx 2$) the magnetization is smeared out by the effective Zeeman term, leading to a significant reduction of the peaks in the magnetization by a factor of 2-3.
 \item In the medium-filling regime we find qualitatively different behavior. Instead of the rapidly oscillating phase we find, depending on the filling, essentially the same structure as is found either in the low- or in the high-filling regime. 
\end{enumerate}

\section{5. Thermodynamically stable vortices}
\begin{figure*}
 \begin{minipage}{0.8\columnwidth}
 \resizebox{0.95\columnwidth}{!}{\includegraphics{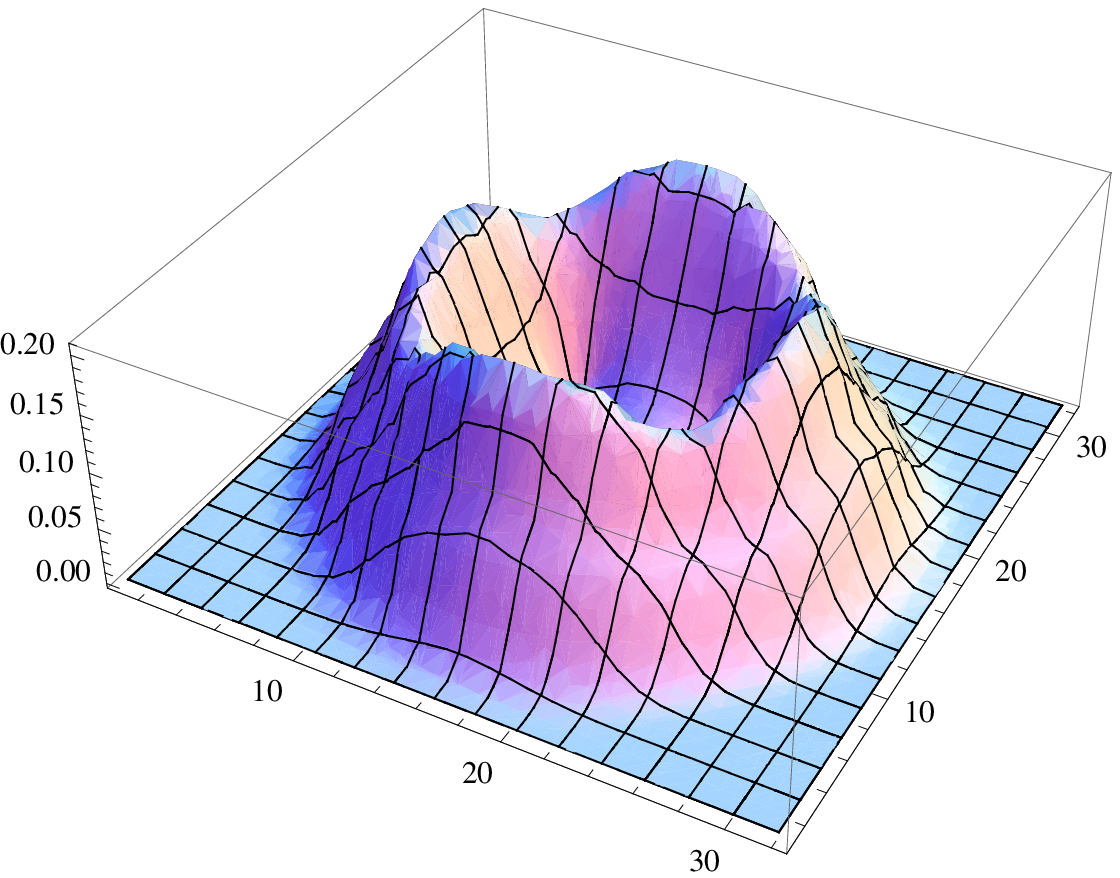}}
 (a)
 \end{minipage}
 \begin{minipage}{0.8\columnwidth}
 \resizebox{0.95\columnwidth}{!}{\includegraphics{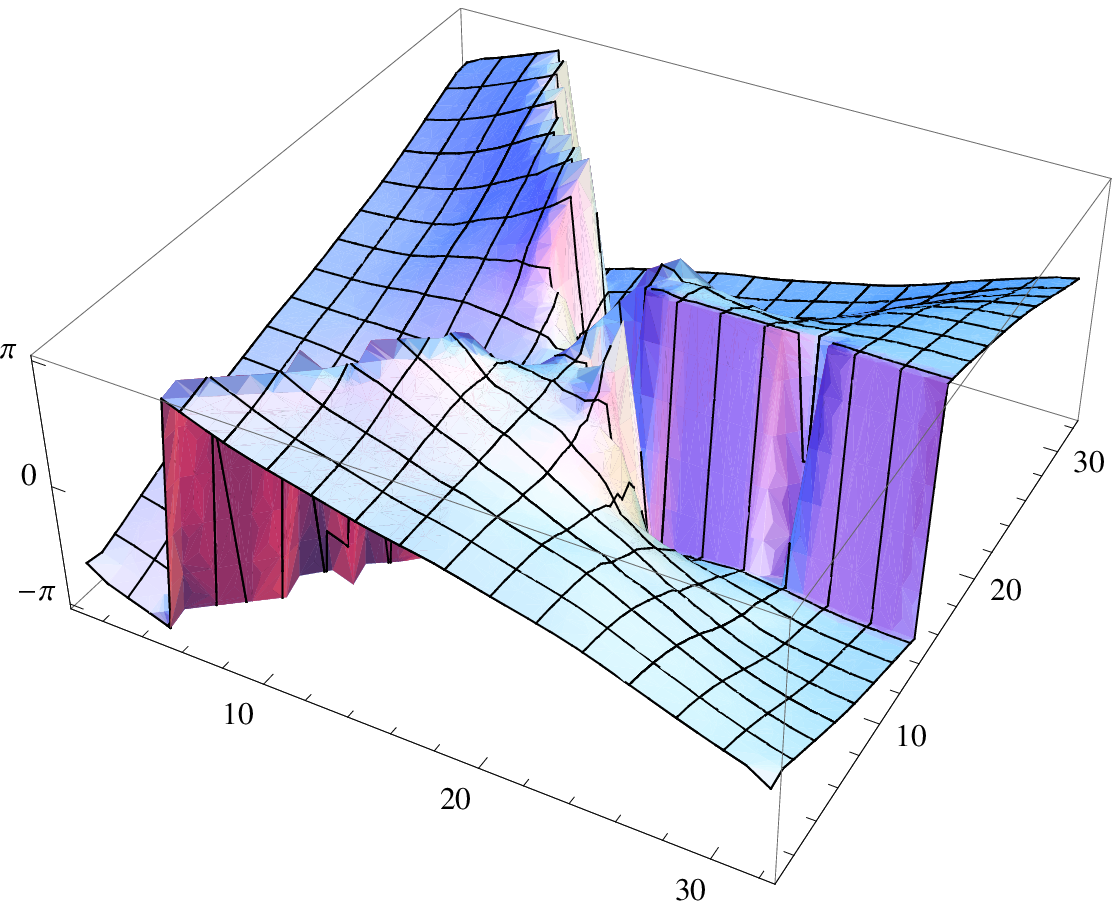}}
 (b)
 \end{minipage}
\caption{$|\Delta(\veci)|$ (a) and $\varphi(\veci)$ (b) in a vortex state with $n=3$ depending on $\veci$-position on a $32 \times 32$ lattice.
The system is balanced and therefore the magnetization is not shown. The system parameters have been chosen as: $U=-2.5$, $V=0.025$, $\beta=1000$, $\mu \equiv (\mu_\uparrow + \mu_\downarrow)/2=0.5$ and $\Delta \mu =0$ in units of $t$.}\label{bgderT}
\end{figure*}

\begin{figure*}
 \begin{minipage}{0.66\columnwidth}
  \resizebox{1.0\columnwidth}{!}{\includegraphics[clip=true]{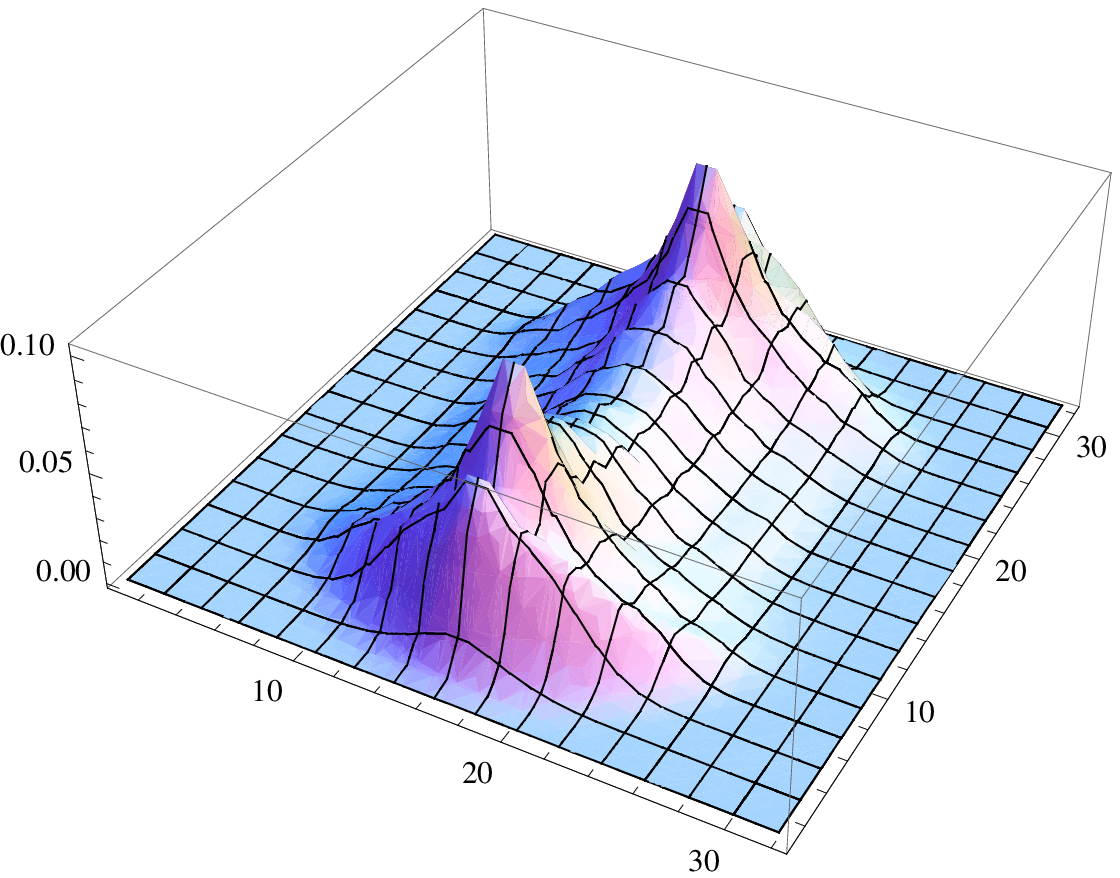}}
 (a)
 \end{minipage}
 \begin{minipage}{0.66\columnwidth}
  \resizebox{1.0\columnwidth}{!}{\includegraphics[clip=true]{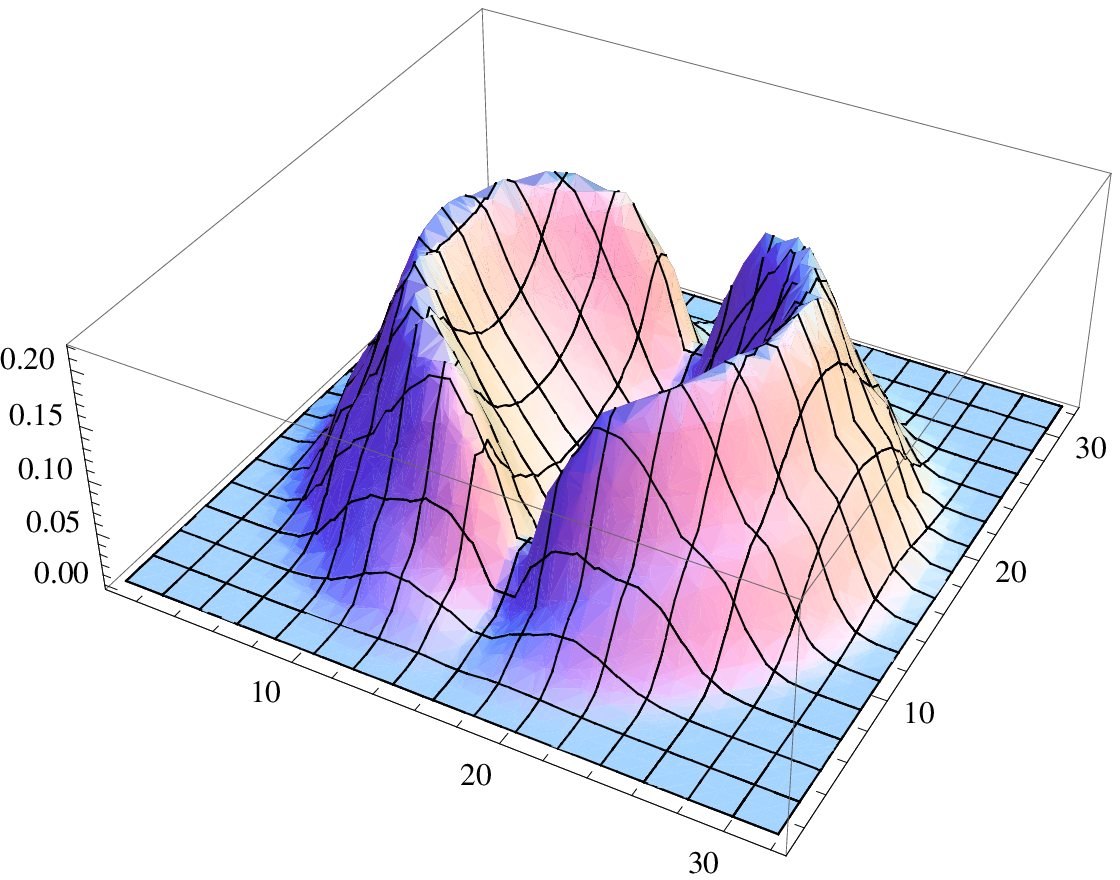}}
 (b)
 \end{minipage}
 \begin{minipage}{0.66\columnwidth}
  \resizebox{1.0\columnwidth}{!}{\includegraphics[clip=true]{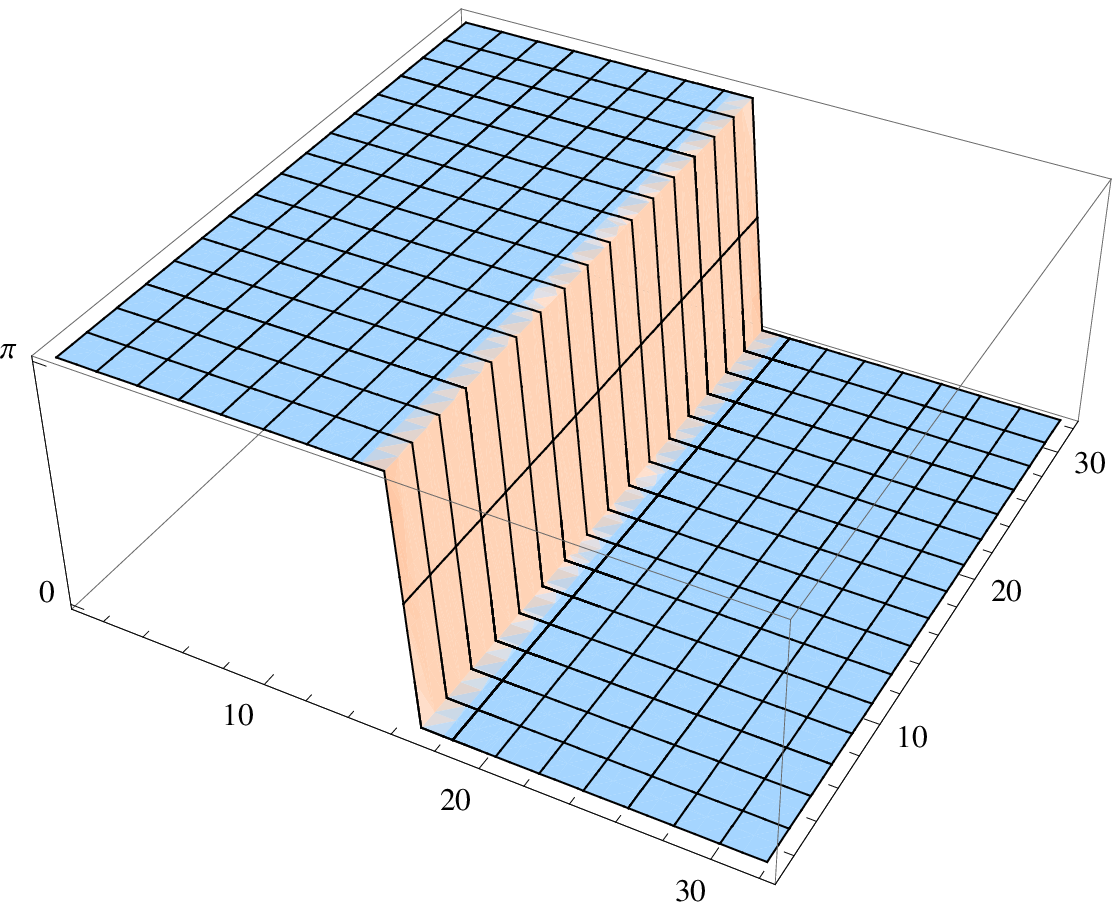}}
 (c)
 \end{minipage}
\caption{Magnetization (a), $|\Delta(\veci)|$ (b) and $\varphi(\veci)$ (c) in the energetically lowest state with the same parameters as in Fig. \ref{gt66ui0}. $\Delta(\veci)$ is a real function. The magnetization breaks the $\pi/2$-rotational symmetry of the lattice.}\label{bkopwW}
\end{figure*}

The derivation of the statement, that the superfluid order parameter can globally be assumed to be a {\it real\/} function, obviously breaks down if the superfluid velocity is singular at any point in real-space, such as happens in the presence of vortices. We will now present numerical results, obtained within the saddle-point approximation, containing such vortices. Note that these solutions do not correspond to the minimum of the grand potential and have, therefore, to be interpreted as excitations, where the superfluid fraction of the system carries a finite angular momentum. Vortices tend to be numerically stable especially in the high-filling regime, in which the order parameter is small in the center of the trap. Results are shown in Figs. \ref{gt66ui0}, \ref{bgderT} and \ref{bkopwW}. In Fig.\ \ref{gt66ui0} we show a solution with vortex-number $n=1$, and, for comparison, we present a configuration lowering the grand potential at the same parameters in Fig.\ \ref{bkopwW}. Note that the magnetization and $\Delta(\veci)$ break the $\pi/2$-rotational lattice symmetry in the energetically lower state, while they do not in the vortex state. In Fig.\ \ref{bgderT} we show a state with vortex-number $n=3$ in a balanced mixture.

These results show that it is possible to obtain numerically stable vortex-solutions in a system with a superimposed strong optical lattice. In contrast to lattice-free systems \cite{zwierlein:fermionic}, the angular motion of the condensate is not free in lattice systems, but is instead caused by nearest-neighbor tunneling processes. In addition, we have been able to determine the dependence of the magnetization distribution on the vortex quantum number $n$.

\section{6. Numerical Results for spin-dependent hopping}
In addition to our analytical results regarding the nature of the superfluid order parameter, we now present some numerical results for a superfluid trapped Fermi-mixture. While we have shown that a real function $\Delta (\vecx)$ is sufficient to describe an FFLO-state, sign changes in $\Delta (\vecx)$ are commonly understood to characterize FFLO-states in a trapped system. Iskin and Williams \cite{iskin:population:0FGw} and Chen et al. \cite{chen:exploring:kl4dw} find these sign changes to be typical for an imbalanced superfluid mixture and interpret them as an FFLO state. Performing a saddle-point approximation \cite{andersen:magnetic:f890,gottwald:antiferro:jhk6} for a modified attractive Hubbard Hamiltonian \eqref{chT6a} with spin-dependent nearest-neighbor hopping we do not find a sign change in $\Delta (\vecx)$ to be typical for an imbalanced mixture. Spin-dependent hopping arises, e.g., if both spin species have different masses \cite{wang:quantum:lf5gtS}
\begin{equation}
 \frac{t_\uparrow}{t_\downarrow} \approx \frac{m_\downarrow}{m_\uparrow} \; .
\end{equation}
If hopping is spin-dependent ($t_\uparrow \neq t_\downarrow$), an imbalanced mixture arises naturally at $\mu_\uparrow = \mu_\downarrow$, since the free bandwidths of both fermion species differ, in contrast to the case of spin-independent hopping ($t_\uparrow = t_\downarrow \equiv t$).

\begin{figure*}
 \begin{minipage}{0.8\columnwidth}
 \resizebox{0.95\columnwidth}{!}{\includegraphics[clip=true]{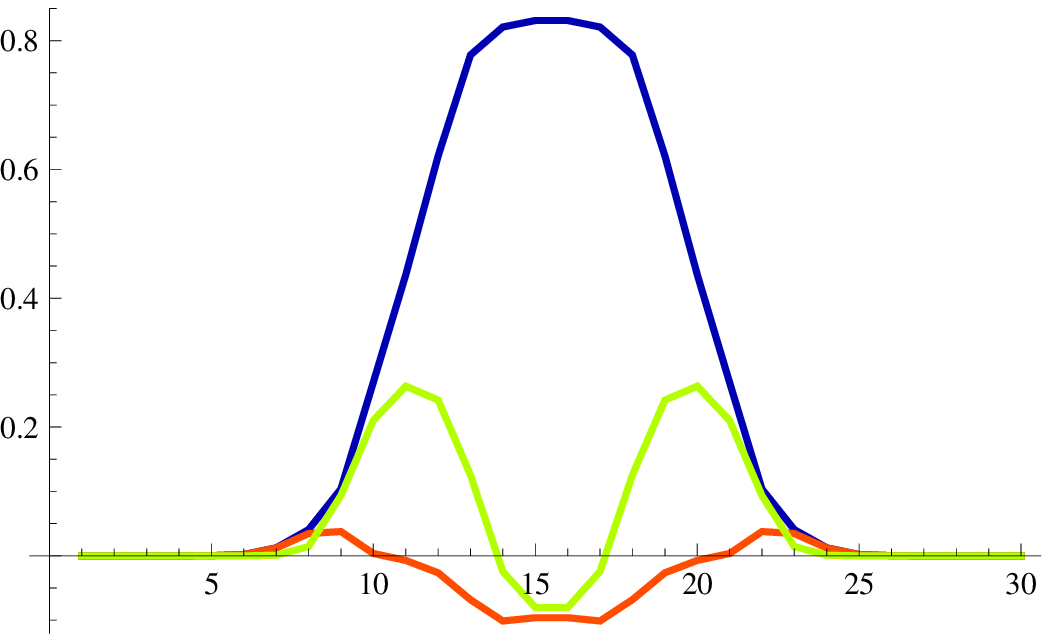}}
  (a)
 \end{minipage}
 \begin{minipage}{0.8\columnwidth}
 \resizebox{0.95\columnwidth}{!}{\includegraphics[clip=true]{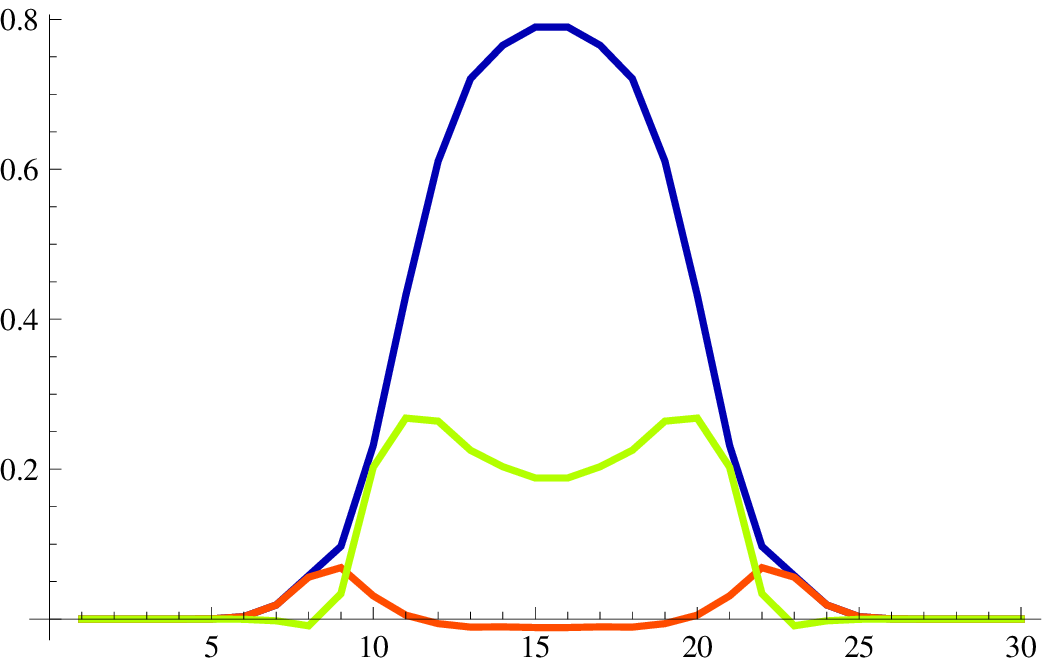}}
 (b)
  \end{minipage}
\caption{Particle density $n \equiv (n_\uparrow + n_\downarrow)/2$ (blue/dark line), population imbalance $m \equiv (n_\uparrow - n_\downarrow)/2$ (orange/grey line) and superfluid order parameter $\Delta$ (green/light grey line) on a $30 \times 30$ square lattice depending on $\veci_x$-position at $\veci_{y}=0$ (slice through the trap center). The parameters are chosen as $U=-2.4$, $V=0.04$, $\beta=50$, $\mu \equiv (\mu_\uparrow + \mu_\downarrow)/2=-0.5$, $t_\uparrow=0.5$ and $t_\downarrow=1$ in both (a) and (b). In the balanced case (a) we have chosen $\Delta \mu = 0.275$ in order to compensate the asymmetry arising from the unequal hopping terms, while in (b) we have $\Delta \mu =0$. The resulting particle numbers are (a) $n_\downarrow \approx n_\uparrow \approx 65$ and (b) $n_\downarrow \approx 56$, $n_\uparrow \approx 71$. $\Delta(\veci)$ has virtually no sign change as a function of real-space in the imbalanced case (b), while the sign changes in the balanced case (a).}\label{bg98Io}
\end{figure*}

Results for superfluid mixtures with asymmetric hopping ($t_\uparrow \neq t_\downarrow$) are shown in Fig.\ \ref{bg98Io}. As one can see, the balanced mixture has a sign-change in the superfluid order parameter, while the imbalanced mixture has virtually no sign change ($\textrm{min} \{\Delta\} \approx -0.009$, while $\textrm{max} \{\Delta\} \approx 0.273$). We conclude, therefore, that, for general hopping amplitudes $t_\sigma$, sign changes in $\Delta (\veci)$ arise from a difference in the chemical potentials $\mu_\downarrow \neq \mu_\uparrow$ and not from an imbalance itself. Furthermore we find that the local magnetization is not generally maximal at sites where the superfluid order parameter has its sign change.

\section{7. Summary and Conclusions}
We investigated the general structure of a $s$-wave pairing order parameter relevant for BCS- and FFLO-states. We have shown that due to the finite size and the absence of vortices in the ground state the superfluid order parameter may always be chosen as a {\it real\/} function, putting into question common complex Ansatzes widely spread in the literature. Furthermore we have investigated the role of the (often neglected) Hartree terms and showed how they qualitatively influence the structure of the superfluid order parameter. Finally we demonstrated that sign changes in the order parameter are not a general feature of particle-imbalanced systems but rather a feature of the grand-canonical parameter $\mu_\uparrow - \mu_\downarrow$, again clarifying and generalizing previous findings.

\bibliography{/home/gottwald/Masterbib/biblography.bib}

\end{document}